\documentclass[a4paper]{article}

\usepackage{INTERSPEECH2021}
\usepackage{multirow}
\usepackage{array}
\usepackage{xcolor}
\usepackage{booktabs}
\newcommand\norm[1]{\left\lVert#1\right\rVert}
\newcolumntype{P}[1]{>{\centering\arraybackslash}p{#1}}
\providecommand{\gz}[1]{\textcolor{black}{{#1}}}

\title{Y-Vector: Multiscale Waveform Encoder for Speaker Embedding}
\name{Ge Zhu$^1$, Fei Jiang$^{1,2}$, and Zhiyao Duan$^1$}
\address{
  $^1$University of Rochester, Rochester, NY, USA\\
  $^2$Beijing Institute of Technology, Beijing, China}
\email{\{ge.zhu, fei.jiang, zhiyao.duan\}@rochester.edu}

\begin{document}

\maketitle
\begin{abstract}
State-of-the-art text-independent speaker verification systems typically use cepstral features or filter bank energies as speech features. 
Recent studies attempted to extract speaker embeddings directly from raw waveforms and have shown competitive results. In this paper, we propose a novel multi-scale waveform encoder that uses three convolution branches with different time scales to compute speech features from the waveform. These features are then processed by squeeze-and-excitation blocks, a multi-level feature aggregator, and a time delayed neural network (TDNN) to compute speaker embedding. We show that the proposed embeddings outperforms existing raw-waveform-based speaker embeddings on speaker verification by a large margin. 
A further analysis of the learned filters shows that the multi-scale encoder attends to different frequency bands at its different scales while resulting in a more flat overall frequency response than any of the single-scale counterparts.
\end{abstract}

\section{Introduction}

In recent years, the development of deep representations of speech utterances has made a breakthrough in speaker verification in terms of accuracy. Variani \textit{et al.}~\cite{VarianiLei14DeepNeural} first trained a deep neural network (DNN) to extract utterance-level features (d-vector), achieving comparable performance to the previous state of the art, i-vector \cite{Dehak11ivector}. Since then, various deep embedding models have been proposed. Among them, x-vector \cite{SnyderRomero18XVector} and its variants \cite{Garcia-Romero2019, Garcia-Romero2020,McCree2019} are the most prominent, achieving state-of-the-art performance in many datasets and tasks \cite{nagrani2020voxsrc}, \cite{Sadjadi2020CTC}.

However, the above-mentioned DNN models are still built upon handcrafted feature inputs such as Mel-Frequency Cepstral Coefficients (MFCCs), which have long been used since Gaussian Mixture Model-Universal Background Models (GMM-UBM). Although MFCCs are designed based on human perceptual evidence, they are not necessarily optimal for speaker recognition tasks and could lose important information during the transform. Thanks to deep learning, there has been a trend on learning feature representations from raw data (e.g., time domain waveforms) to breakthrough the limit of feature engineering~\cite{palaz2019end, Muckenhirn2019}. 

Speaker verification research also witnessed an increased effort on developing time-domain deep neural network approaches.
Taking raw waveforms as the input, a 1-d convolutional layer is usually applied as the first layer, where the set of filters behave like the Short-Time Fourier Transform (STFT), resulting in time-varying filter responses for future layers to process. In \cite{Muckenhirn2018}, Muckenhirn \textit{et al.}~first applied a Convolution Neural Network (CNN) based architecture for speaker verification and achieved competitive results to i-vector on Voxforge dataset\footnote{An open source speech database: http://www.voxforge.org/}. By analyzing the frequency response of the learned filters, they found that the first layer of the CNN was able to implicitly model the fundamental frequency ($F_0$). To efficiently learn meaningful filters, Ravanelli and Bengio proposed SincNet \cite{Ravanelli2018, ravanelli2018interpretable} to constrain the free filters in the first convolutional layer with parameterized sinc functions. Jung \textit{et al.} \cite{jung2020improved} later utilized this sinc-convolution layer with RawNet \cite{jung2019RawNet} and feature map scaling, and marginally outperformed the existing  best spectrogram based system. In \cite{Lin2020Wav2SpkAS}, Lin and Mak adapted the architecture in wav2vec \cite{schneider2019wav2vec} and achieved an equal error rate (EER) of 1.95$\%$ on the VoxCeleb1-O test set.

However, the set of filters in a convolutional layer typically has the same kernel size. This makes it difficult to learn high-frequency and low-frequency components simultaneously for wide-band signals. One idea is to split one convolution branch into several parallel branches with different scales, similar to InceptionNet \cite{Szegedy2015} in computer vision. In this way, different groups of parameters for the convolution layer, including the number of filters, kernel size and stride size, can be independently determined, and filters at each scale can respond to different frequency components efficiently. Multi-scale convolutions have also been successfully used in acoustic modeling for speech recognition tasks from the raw waveform \cite{Zhu2016LearningMF,Platen2019MultiSpanAM}. This also motivates us to learn time-domain multi-scale representations for speaker verification.

In this paper, we present a new time-domain speaker embedding (Y-vector) based on a novel multi-scale waveform encoder. Compared to existing time-domain approaches \cite{Lin2020Wav2SpkAS}, the proposed system uses a multi-scale waveform encoder to capture broadband responses. It also uses a time-frequency squeeze-excitation ($tf$-SE) attention module to re-calibrate the importance across time and frequency domains, and a TDNN for frame aggregation. Extensive experiments are conducted on the VoxCeleb1-O, VoxCeleb1-H and VoxCeleb1-E test sets. Results show that Y-vector outperforms existing time-domain speaker verification systems by a large margin. 
Further analysis shows that the multi-scale encoder responds to different frequency bands at its different scales, while resulting in a more flat overall frequency response than its single-scale counterparts.

\section{Proposed System}

\label{sec:ProposedSystem}
\begin{figure}[!t]
\centering
\includegraphics[width=2.4in]{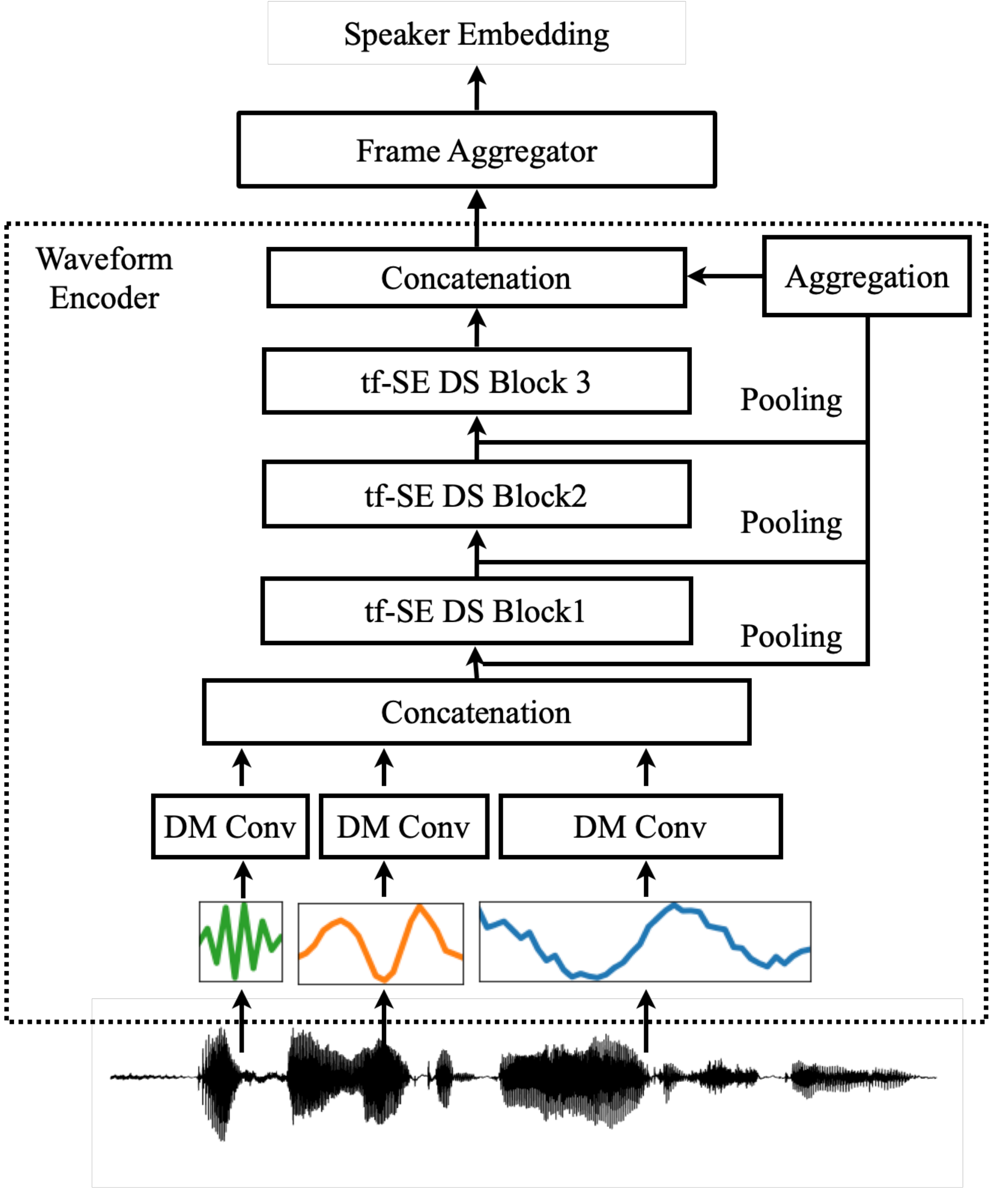}
\caption{Block diagram of our proposed Y-vector system. In the waveform encoder part, the three wave shaped curves demonstrate learned filters at different temporal resolutions. (DM: dimension match, DS: downsampling)} 
\label{fig:diagram}
\end{figure}

The proposed Y-vector system is shown in Fig.~\ref{fig:diagram}. First, the multi-scale waveform encoder uses two filtering layers to take the same raw waveform input into multiple streams operating at different temporal resolutions. The filtered embeddings are then concatenated and then go through three ($tf$-SE) convolutional downsampling blocks. Finally, this representation is fed into a frame aggregator, implemented as a TDNN with additive margin softmax (AM-Softmax) to extract speaker embeddings using a speaker classification task. We now describe the details of each stage.
\subsection{Multi-scale Filtering Layer}


Different from STFT that uses analytical Fourier basis as the filters, raw waveform encoders learn convolution filters to process the waveform. They need to use a small stride size, because a large size would lead to low temporal resolution and the loss of important information. Therefore, in \cite{Zhu2016LearningMF}, Zhu \textit{et al.} used a small stride convolutional layer followed by max pooling. 
Similarly, in wav2vec \cite{schneider2019wav2vec}, Schneider \textit{et al.} applied a five-convolutional-layer encoder with a series of relatively small strides of \{5, 4, 2, 2, 2\}, to decrease the time dimension gradually. With this strategy, the final sequence length of  audio samples becomes 160 times smaller than the original input, while important information can passes through the layers more easily for learning good representations for speakers \cite{Lin2020Wav2SpkAS}.

However, the above methods both use a single-scale convolution filterbank to process the waveform, which limits the frequency responses to the input speech signal. To be specific, short filters do not have the sufficient length to respond to low frequencies, while long filters can be inaccurate in modeling high frequencies as the signal might be non-stationary within the filter window. 

In this work, we propose to extend the encoder into a multi-scale setting. To do so, we replace the first two convolution layers of wav2vec with three parallel branches at different scales, which are reflected by the filter size. Each branch consists of two layers of 1-d convolution, shown in the lower part of Fig.\ref{fig:diagram}. The first layer is used for primary filtering and the second dimension-match layer aims to compensate for the resulted output dimension differences. Therefore, the multiplication of the stride sizes is a constant across the three branches. It is also possible to use more than two layers, but in this paper, we only investigate the two-layer case. 

\subsection{$tf$-SE Downsampling Block}
After concatenating the feature maps from different branches, we use three convolution blocks to further downsample them into a feasible size for the subsequent frame aggregator. The architecture of the downsampling block can be written as:
\begin{equation}
    Y = tf\text{-SE}(\text{ReLU}( \text{Norm}(\text{Dropout}(\text{Conv}(X))),
\end{equation}
where $X,Y\in \mathcal{R}^{F \times T}$ are the input and output feature maps, respectively, $F$ is the number of filters and $T$ is the sequence length in time, $tf$-SE denotes for the temporal and frequency squeeze and excitation module. 
This idea is borrowed from the attention module for 2D signal processing~\cite{Hu18SE, yadav20ft}, where the importance of different channels of the embeddings is re-calibrated through squeeze and excitation networks. In our work, the re-calibration is performed in time and frequency dimensions instead, in a sequential manner. Specifically, we first aggregate the global information of the whole utterance from the input using average pooling along time, then re-scale the frequency dimension through:
\begin{equation}
    X'=\sigma(\mathbf{W}_1 AvgPool_{1:T}(X)+\mathbf{b}_1)\odot X,
\end{equation}
where $\mathbf{W}_1$ is a matrix of size $F \times F$, $\mathbf{b}_1$ is a bias vector of size $F \times 1$, $\odot$ denotes for element-wise multiplication and $\sigma$ denotes for sigmoid function. During time re-calibration, we borrow the idea of temporal gating in \cite{Lin2020Wav2SpkAS}, re-scaling feature maps at every time frame $t$ with a scalar factor. Then the resulting output feature at frame $t$ can be written as:
\begin{equation}
    Y_t=\sigma(\mathbf{W}_2 (X_t')+\mathbf{b}_2))\odot X_t',
\end{equation}
where $\mathbf{W}_2$ is a matrix of size $F \times 1$ and $\mathbf{b}_2$ is a bias scalar. \gz{By collecting the global ``time" information in the first step, all of the ``frequency" components will be re-calibrated with excitation module. Similarly, the second step performs as a gate mechanism, all of the ``time" frames will be re-weighted.}

\subsection{Multi-level Feature Map Aggregation}
The original wav2vec only uses the feature map of the last layer for further processing. Although deeper layer features are usually more complex and contribute more to the final representations, Lee \textit{et al.} \cite{Lee2017MultiLevel} found that features extracted by early layers are also helpful through skip connections. Therefore, to further improve the accuracy, we concatenate feature maps at different layers at each time frame. To do so, max pooling is used to downsample earlier-layer feature maps to the same frame rate as that of the last layer.

\section{Experiments}
\label{sec:typestyle}
\subsection{Dataset}
In our experiment, we use the VoxCeleb corpora \cite{Chung18b, Nagrani17VoxCeleb1} to train and evaluate our system and comparison systems for speaker verification. VoxCeleb is a free large-scale  text-independent dataset collected from public multimedia data, where acoustic conditions are not controlled. Specifically, we employ the VoxCeleb2 development dataset for training, which contains 2,442 hours of recordings from 5,994 speakers. VoxCeleb1 is used for testing. It contains three test sets: (i) VoxCeleb1-O: the original verification test set consists of 37,611 random pairs from 40 speakers; (ii) VoxCeleb1-E: a list of 579,818 random pairs; (iii) VoxCeleb1-H: a list of 550,894 pairs with the same nationality and gender. Among these test sets, VoxCeleb1-E and VoxCeleb1-H cover all the 1251 speakers in VoxCeleb1. 

For the evaluation metric, we compute Equal Error Rate (EER) and the minimum of the normalized detection cost function (minDCF) at $C_{miss}=C_{fa}=1$ and $P_{Target}=10^{-2}$ on these test sets to measure speaker verification accuracy. At test phase, we apply cosine score backend to all systems to measure the similarity between testing pair embeddings and calculate EER by adjusting the decision threshold.

\subsection{Multiscale Architecture}
The detailed multi-scale waveform encoder of Y-vector is shown in Table \ref{tab:encoder}. Here we fix the ratio between stride size and kernel size to 0.5, which is equivalent to a 50$\%$ overlap ratio in STFT. In our study, we use a TDNN \cite{SnyderRomero18XVector} as the frame aggregator.
\begin{table}[ht!]
\setlength{\tabcolsep}{4pt}
  \centering
  \caption{System Y-vector-5. Numbers in brackets are convolution parameters: number of channels, kernel size and stride size, respectively.}
  \scalebox{0.85}{
  \begin{tabular}{P{80pt}|P{40pt}|P{40pt}|P{40pt}}
    \toprule
    \textbf{Group}    &\multicolumn{3}{c}{\textbf{Conv. Parameters}}\\
    \hline
    \multirow{3}{*}{Multi-scale Filtering} &Branch 1&Branch 2&Branch 3\\
     \cline{2-4}
    &$\left[90, 12, ~6\right]$&$\left[90,18,9\right]$&$\left[90,36,18\right]$ \\ 
    &$\left[160,5,3\right]$&$\left[160,5,2\right]$&$\left[192,5,1\right]$ \\
    \hline
      Concatenation  &\multicolumn{3}{c}{-}\\
    \hline
    \multirow{3}{*}{Downsampling} &\multicolumn{3}{c}{$\left[ 512,5,2 \right]$}\\
    &\multicolumn{3}{c}{$\left[ 512,3,2 \right]$} \\ &\multicolumn{3}{c}{$\left[ 512,3,2 \right]$} \\    
    \bottomrule
  \end{tabular}}
  \label{tab:encoder}
\end{table}

For the ablation study in Section \ref{sssection:Ablation}, we design several variants of the proposed system. Specifically, we compare systems with different number of channels in the first layer, as it can be viewed as the ``frequency resolution" counterpart in STFT. We also compare systems with different total decimation rates of the first two layers, i.e., the multiplication of their stride sizes, as this indicates how fast the time dimension is reduced.
We also investigate the effectiveness of multi-level aggregation and $tf$-SE components. Details are listed in Table~\ref{tab:mwsystems}. 
\begin{table}[ht!]
\small
\setlength{\tabcolsep}{4pt}
  \centering
  \caption{Different multi-scale waveform encoder variants explored in ablation study.}
  \scalebox{0.8}{
  \begin{tabular}{P{40pt}|P{40pt}|P{42pt}|P{35pt}|P{30pt}}
    \toprule
    \textbf{System}    & \textbf{\# of Channels}& \textbf{Decimation Rate}& \textbf{ML Aggreg.} & \textbf{$tf$-SE.}\\
    \hline
    Y-vector-1& 150 & 24 &   & \\
    \hline
    Y-vector-2& 150 & 24 & \checkmark & \\
     \hline
    Y-vector-3& 270 & 24 & \checkmark & \\
     \hline
    Y-vector-4& 270 & 18 & \checkmark & \\
    \hline
    Y-vector-5& 270 & 18 & \checkmark & \checkmark\\
    \bottomrule
  \end{tabular}
  }
  \label{tab:mwsystems}
\end{table}
\subsection{Training Details}
\begin{table*}[!t]
\caption{EER (\%) comparison on different test sets. All models are trained on the VoxCeleb2 training set and scored with cosine similarity. A statistical significance test is performed using a bootstrap procedure~\cite{haasnoot18eerci}: Because Vox1-E and Vox1-H are much larger than Vox1-O, an absolute value of 0.05 of EER difference for Vox1-E and Vox1-H is already outside the 95$\%$ confidence interval for all methods, while for Vox1-O the EER difference has to be larger than 0.15. ($\ast$: results copied from the references. $\dagger$: our implementation. SP: statistical pooling.)} 
\renewcommand{\arraystretch}{1}
\small
\setlength{\tabcolsep}{4pt}
  \centering
\scalebox{0.9}{
\begin{tabular}{P{100pt}P{45pt}P{50pt}P{47pt}P{25pt}P{30pt}P{25pt}P{30pt}P{25pt}P{30pt}}
\toprule
\multirow{2}{*}{\textbf{Method}} & \multirow{2}{*}{\textbf{Feature}} & \multirow{2}{*}{\textbf{Aggregation}} &\multirow{2}{*}{\textbf{Loss}} &\multicolumn{2}{c}{\textbf{VoxCeleb1-O}}& \multicolumn{2}{c}{\textbf{VoxCeleb1-E}} & \multicolumn{2}{c}{\textbf{VoxCeleb1-H}} \\ 
\cline{5-10} \specialrule{0em}{1pt}{1pt} 
&&&&EER&minDCF&EER&minDCF&EER&minDCF\\
\toprule
 Monteiro et al. \cite{monteiro20e2e} &MFCC & SP&E2E & 2.51$^\ast$&-& 2.53$^\ast$&-&4.69$^\ast$&-  \\
3 CNN+x-vector$^\dagger$ &MFCC & SP&AM-softmax  &2.82 &0.284 &2.96&0.302&4.91&0.422 \\
\hline
Xie et al. \cite{Xie19} & \multirow{2}{*}{Spectrogram} & GhostVLAD&Softmax& 3.24$^\ast$&-& 3.13$^\ast$ &-& 5.06$^\ast$&- \\
Nagrani et al. \cite{Nagrani19} &  & GhostVLAD&Softmax& 2.87$^\ast$& 0.310$^\ast$& 2.95$^\ast$ &- & 4.93$^\ast$&- \\
\hline
RawNet2 \cite{jung2020improved} & \multirow{8}{*}{Raw Waveform} & GRU&Softmax& \textbf{2.48}$^\ast$&-& 2.87$^\ast$ &-& 4.89$^\ast$&- \\
wav2spk$^\dagger$ \cite{Lin2020Wav2SpkAS} & & Gating + SP &AM-softmax & 3.00&0.281 & 2.78&0.280&4.56&0.390 \\
\textbf{ modified wav2spk} & & SP&AM-softmax &2.69&0.278  &2.62&0.261& 4.28&0.371 \\
\textbf{Y-vector-1} \textbf{(ours)}& & SP  &AM-softmax & 2.78 &0.269 &2.64 &0.270& 4.33& 0.377\\
\textbf{Y-vector-2} \textbf{(ours)}& & SP  &AM-softmax & 2.77 &0.270&2.50 &0.263& 4.17&0.376\\
\textbf{Y-vector-3} \textbf{(ours)}& & SP  &AM-softmax & 2.79 &0.258 &2.47 &0.256& 4.07& 0.366\\
\textbf{Y-vector-4} \textbf{(ours)}& & SP  &AM-softmax & 2.60 &0.239 & 2.39 &0.248& 4.00& 0.354\\
\textbf{Y-vector-5} \textbf{(ours)}& & SP  &AM-softmax & 2.72 &0.261 & \textbf{2.38} &\textbf{0.241}& \textbf{3.87}& \textbf{0.339}\\
\bottomrule
\end{tabular}
}
\label{tab:results}
\end{table*}
At the preprocessing stage, we simply normalize the raw waveform of each utterance by its maximum value. No voice activity detection (VAD) module is used. All of the recordings from VoxCeleb2 are used without filtering out speakers with short utterances, and we did not perform any data augmentation tricks either. For each utterance, we randomly crop 3.9s for batchifying to feed to the neural network.

For the TDNN frame aggregator, we empirically find that layer normalization works better than batch normalization in our system. We also apply L2 regularization on the last two fully connected layers combined with LeakyReLU activation functions with a negative slope of 0.2, following the method mentioned in \cite{Zeinali19}. As for the AM-Softmax loss function, the scale factor and margin are set to 30 and 0.35 respectively. For training, we use Stochastic Gradient Descent (SGD) with an initial learning rate of 0.01 and a momentum of 0.9. The learning rate decays by a factor of 0.5 for every 60 epochs. We train the system for 300 epochs, and in each epoch, we randomly sample 240,000 utterances from the whole training set. The batch size is set to 96. 

\subsection{Results}
\label{Results}

\subsubsection{Comparison with Other Systems}

In this section, we compare the proposed Y-vector with other recent speaker embedding networks using various features. The results are listed in Table \ref{tab:results}. It can be seen that Y-vector significantly outperforms all other raw waveform-based systems and spectrogram-based systems on both VoxCeleb1-E and VoxCeleb1-H. It is noted that one comparison method, modified-wav2spk, is our implementation of wav2spk with the proposed multi-scale encoder using the same number of input channels; we also remove the original temproal gating because it already appears in $tf$-SE modules. Comparing with an MFCC-based system \cite{monteiro20e2e} with a similar backbone neural architecture, we can see that Y-vector also achieves better performance.  One might argue that this difference might be due to the complexity of the 5-layer convolution waveform encoder in Y-vector instead of the benefit of raw waveform input. To verify this, we build another system (`3 CNN + x-vector') that takes MFCC as input, and feeds it to the last three convolution layers of the waveform encoder followed by the TDNN aggregator. We used a 3-layer CNN because it has been shown in \cite{anden14deepscattering} that standard mel-filterbanks can be approximated by two convolution operations. However, as can be seen in Table \ref{tab:results}, the MFCC system still underperforms Y-vector systems significantly. \gz{Notice that, our proposed waveform system still cannot compete with state-of-the-art spectrum-based systems \cite{Brecht20ecapa} due to different experimental settings, model architectures and other tricks, which is beyond the scope of this paper.}

\subsubsection{Ablation Study of the Y-vector Architecture}
\label{sssection:Ablation}
In this section, we compare five variants of the proposed Y-vector system listed in Table \ref{tab:mwsystems}. From Y-vector-1 to Y-vector-2, we see improvement on both VoxCeleb1-E and VoxCeleb1-H, showing the effectiveness of multi-level feature aggregation. When increasing frequency resolution of filters (from Y-vector-2 to Y-vector-3) and decreasing total decimation rate (from Y-vector-3 to Y-vector-4), the performance both improves. Finally, when comparing Y-vector-4 with Y-vector-5, we see that although the $tf$-SE block does not improve the performance in VoxCeleb-E, it does improve on VoxCeleb-H, which is comprised of \textit{harder} trials with more similar utterances in each trial.

\subsubsection{Multi-scale Versus Single-scale}
We also compare the multi-scale waveform encoder with three different single-scale encoders. We use a filter size of 10, 20, 40 respectively in the three single-scale encoders with each contains 96 channels. For the multi-scale encoder, we use the above three filter sizes in parallel branches to model high, middle, and low frequency information, respectively. For a fair comparison, we use only 32 filters in each branch so that the total number of filters is equal to that of the single-scale encoders, which means that the number of parameters is nearly the same.

\begin{figure}[!h]
\centering
\includegraphics[width=2.1in]{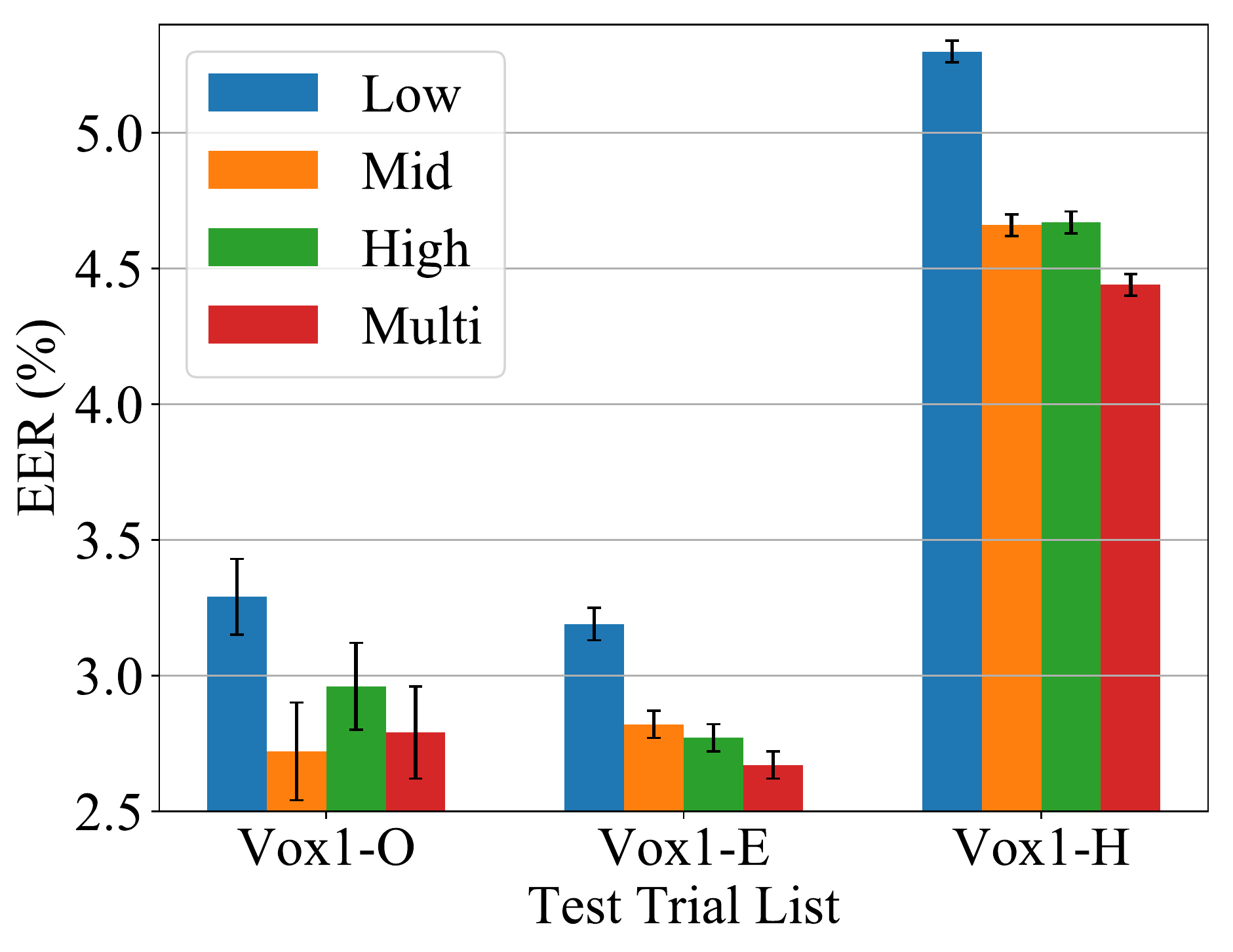}
\caption{Test set EER ($\%$) comparing single- and multi-scale versions of the proposed system. Error bars show a 95$\%$ confidence interval.}
\label{fig:scaleeer}
\end{figure}

The evaluation results of the four encoders are shown in Fig.\ref{fig:scaleeer}. We can see that the ``low" encoder performs the worst, ``mid" and ``high" encoders perform similarly with each other, while the “multi” encoder improves the EER over other systems on Vox1-E and Vox1-H slightly.\\

\begin{figure}[!t]
\centering
\includegraphics[width=2.5in]{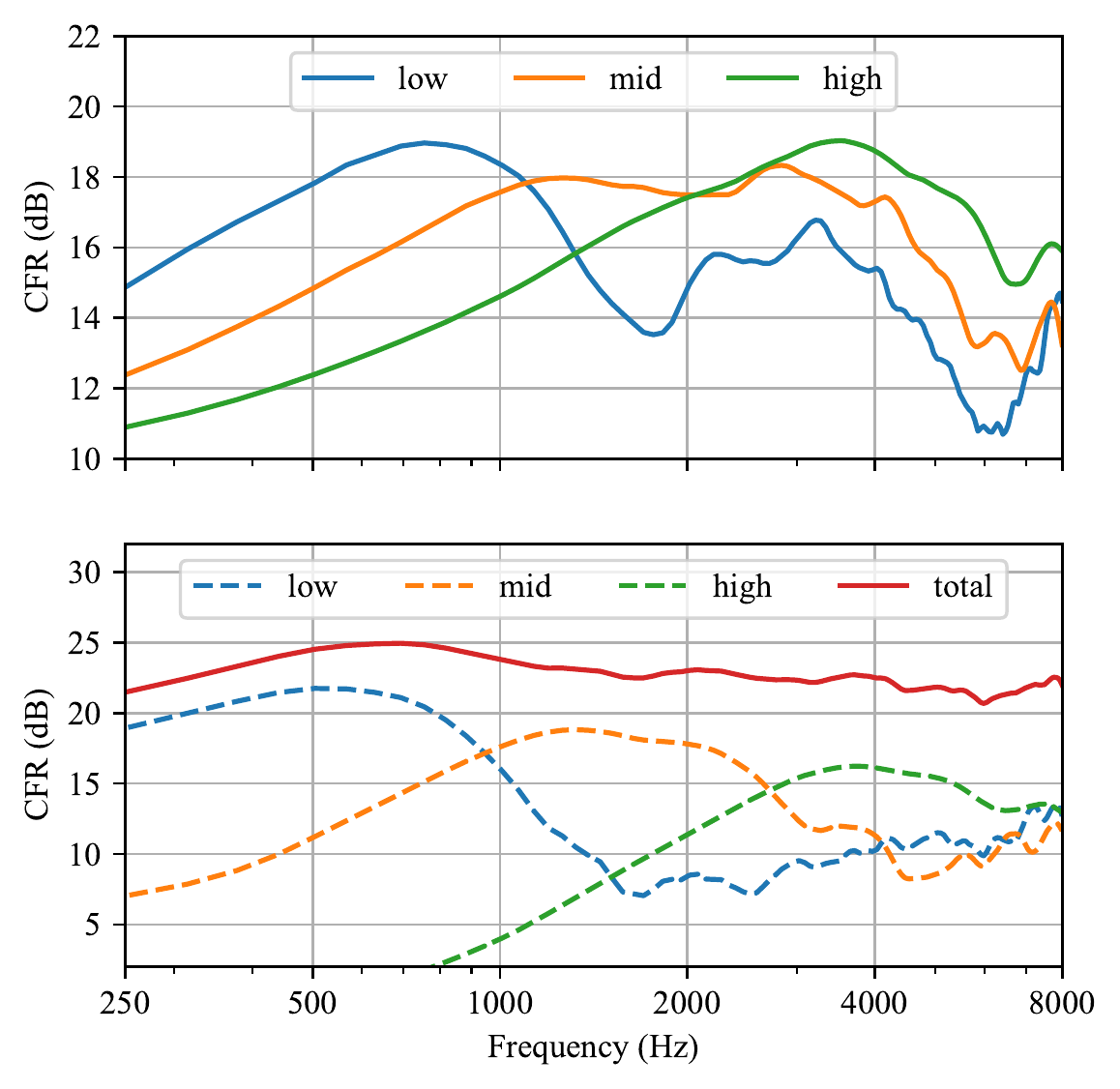}
\caption{Top: CFRs of the learned filters at different scales of the single-scale waveform encoder. Bottom: CFRs of the learned filters of the multi-scale waveform encoder and its single-scale branch.} 
\label{fig:cfr}
\end{figure}

To further investigate the attributes of the multi-scale filters in the encoder, we compute the Cumulative Frequency Response (CFR) of the learned filters in the previous section as \cite{palaz2019end}:
\begin{equation}
    CFR=\sum_{k=1}^{M}\frac{\mathcal{F}_k}{\norm{\mathcal{F}_k}_2},
\end{equation}
where $\mathcal{F}_k$ is the magnitude frequency response of the filter $f_k$ computed with a 256-point discrete Fourier transform, and $M$ is the number of filters. The CFRs of the learned filters in the three single-scale and the multi-scale waveform encoders are shown in the top and bottom of Fig. \ref{fig:cfr}, respectively. As shown in the top figure, the three single-scale encoders focus on different frequency bands but none of them is capable of modeling a wide frequency range. In contrast, the bottom figures shows that the multi-scale encoder has a much more flat overall CFR covering the entire frequency range with less than 5 dB fluctuations. This owes to the filters in the three parallel branches responding to different frequency bands. 
Note that the overall CFR has a peak between 400 Hz and 1000 Hz, suggesting that this frequency band plays a more important role in speaker recognition.

\section{Conclusion}

In this paper, we proposed a multi-scale raw waveform speaker embedding system and demonstrated its effectiveness in speaker verification. To be specific, the multi-scale waveform encoder uses three convolution branches with different time scales to compute speech features from the waveform, which are then processed by squeeze-and-excitation blocks and a multi-level feature aggregator. On speaker verification, the proposed system significantly outperforms both time-domain and several MFCC-based speaker embedding systems, on both VoxCeleb1-H and VoxCeleb1-E. Future work includes the investigation of learnable filter bank architectures and cross-domain tasks especially under channel mismatch conditions.
\section{Acknowledgements}
This work was supported by National Science Foundation grant No. 1741472 and funding from Voice Biometrics Group.
\bibliographystyle{IEEEtran}

\bibliography{mybib}
\end{document}